# On the effect of elastic nonlinearity on aquatic propulsion caused by propagating flexural waves

V.V. Krylov

Department of Aeronautical and Automotive Engineering,
Loughborough University,
Loughborough, Leicestershire, LE11 3TU, UK

## Abstract

In the present paper, the initial theoretical results on wave-like aquatic propulsion of marine craft by propagating flexural waves are reported. Recent experimental investigations of small model boats propelled by propagating flexural waves carried out by the present author and his co-workers demonstrated viability of this type of propulsion as an alternative to a well-known screw propeller. In the attempts of theoretical explanation of the obtained experimental results using Lighthill's theory of fish locomotion, it was found that this theory predicts zero thrust for such model boats, which is in contradiction with the results of the experiments. One should note in this connection that Lighthill’s theory of fish locomotion assumes that the amplitudes of propulsive waves created by fish body motion grow from zero on the front (at fish heads) to their maximum values at the tails. This is consistent with fish body motion in nature, but is not compatible with the behaviour of localised flexural waves used for propulsion in the experimental model vessels. To overcome the problem of predicting zero thrust for the above vessels, it is suggested in the present work that nonlinear distortion of localised flexural waves may be responsible for generation of thrust. This hypothesis is explored in the present work by adding nonlinear harmonics of propulsive flexural waves to the Lighthill's formula for generated thrust. For simplicity, only the lowest (third) harmonic growing linearly with the distance of propagation was used. The resulting formula for the averaged thrust shows that, due to the effect of the third harmonic, the thrust is no longer zero, thereby demonstrating that nonlinear distortion of the propulsive flexural waves may be paramount for the existence of wave-like aquatic propulsion of small marine craft using freely propagating flexural waves.

## 1 Introduction

The most common method of aquatic propulsion used in existing marine vessels is a screw propeller. However, the conventional propeller has a number of disadvantages. In particular, these are cavitation and generation of the associated underwater noise. One of the possible ways to overcome these and some other problems associated with a propeller is to develop alternative propulsive systems, in particular systems inspired by nature and simulating fish swimming by employing flexural elastic wave propagation in immersed plate-like structures.

The idea of wave-like aquatic propulsion of manned marine vessels to be dealt with in the present paper has been first published in 1994 by the present author [1] (see also the paper [2]). This idea is based on using the unique type of localised flexural elastic waves freely propagating along edges of slender wedge-like structures immersed in water. Such wedge-like structures supporting the above-mentioned localised elastic waves, also known as wedge elastic waves, can be attached to a body of a small ship or a submarine as keels or wings that are to be used for aquatic propulsion (see Fig.1). The above-mentioned wedge elastic waves freely propagating in contact with water have been first predicted and investigated theoretically in the above-mentioned papers [1, 2]. Note that the principle of using these wedge elastic waves as a source of aquatic propulsion is similar to that used in nature by stingrays, which utilise wave-like motions of their large horizontal pectoral fins (wings) for moving forward.

It is vitally important for the application of localised elastic waves for propulsion of manned marine vessels that, in spite of vibration of the attached fins, the main body of the craft remains undisturbed because the energy of localised waves is concentrated near the wings' tips [1]. In comparison with a propeller, the above-described wave-like aquatic propulsion has the following main advantages: it does not generate cavitation and the associated underwater noise; it is efficient, approaching the efficiency of dolphins and sharks; and it is safe for people and marine animals.

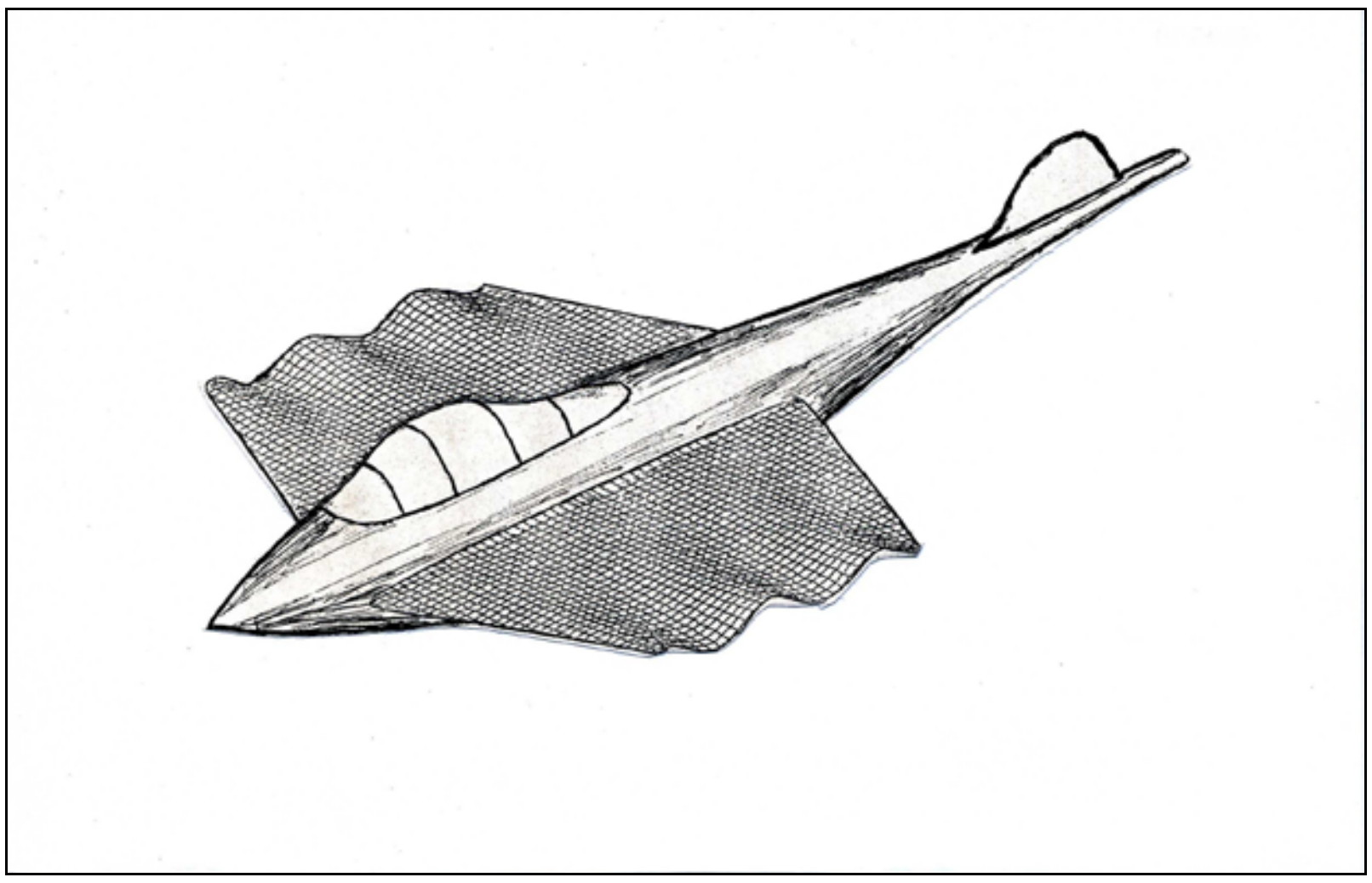

Fig. 1 Illustration of the proposed use of localised flexural waves propagating along tips of wedge-like structures, also known as wedge elastic waves, for propulsion of a small submarine [1].

The first practical implementations and experimental testing of this type of aquatic propulsion have been carried out using a small model catamaran employing localised flexural waves propagating in a vertical rubber plate [3, 4] and a mono-hull model boat propelled by a localised flexural wave propagating along its keel [5]. The test results have shown that both model vessels (the catamaran and the mono-hull boat) were propelled quite efficiently, thus demonstrating that the idea of wave-like aquatic propulsion of manned marine craft is viable. Note in this connection that earlier designs of wave-like aquatic propulsion that were using 'usual' (non-localised) flexural waves [6, 7] caused craft body to be rocking in response to plate vibrations. Therefore, these designs were unsuitable for manned marine craft.

In the present paper, we report some initial results on the theory of wave-like aquatic propulsion by propagating flexural waves that could be applied to small marine craft experimentally tested in [3-5]. It was natural to try to apply the well-known Lighthill's theory of fish locomotion [8, 9] to the above-mentioned model marine craft in the attempts to fit the theory to the experiments [3-5]. However, it was found that Lighthill's theory in its standard form does not give meaningful results for these experiments as it predicts zero thrust at all frequencies due to the particular design of the experimental craft propulsion mechanisms. Namely, the localised flexural waves used for propulsion in [3-5] were excited from the front edges of the propulsive plates via mechanical arms (Fig 2), which means that the amplitudes of these waves were maximal at the front edges of the propulsive plates and were decreasing (or at best were kept constant) over the length of the plates towards the rear edges. According to Lighthill's formula [8], this results in zero thrust, which is in contradiction with the experiments of [3-5], in particular with the measured nonzero values of generated thrust (see Fig. 3).

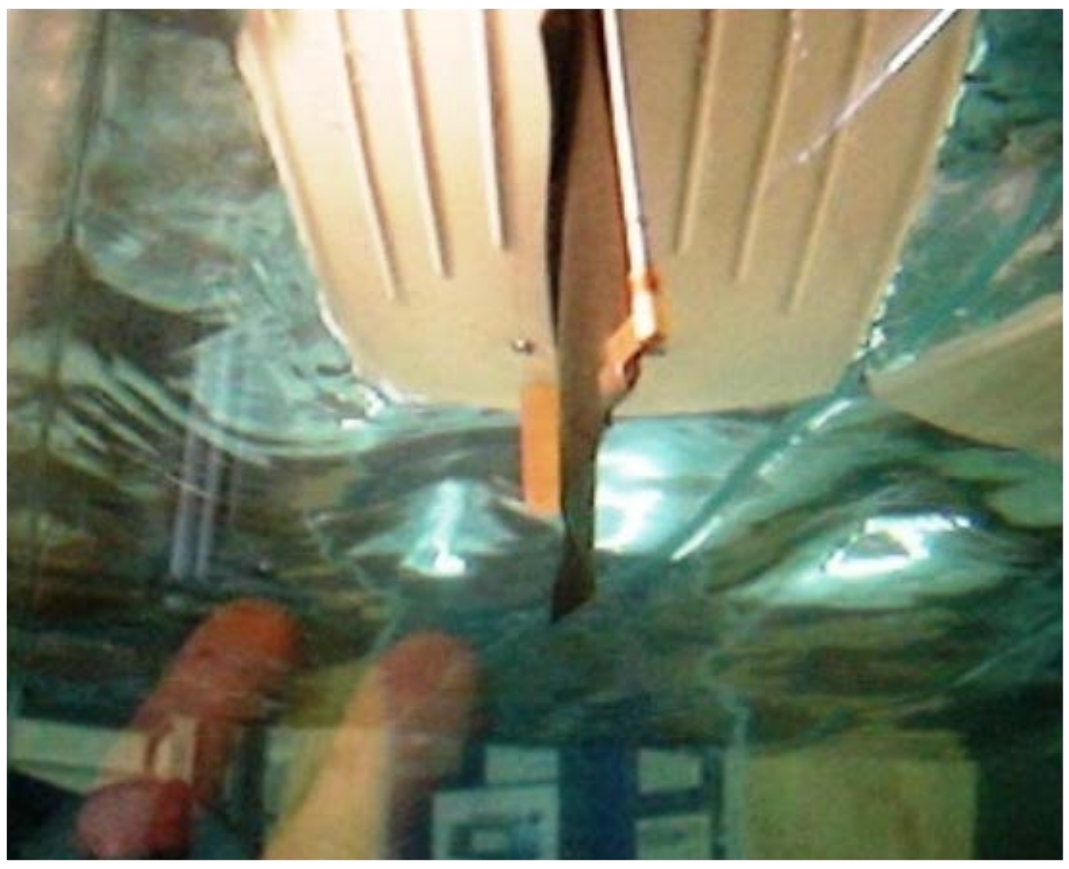

Fig. 2 Underwater view of the mono-hull model boat with the assembled propulsive rubber plate showing the propagation of a localised flexural wave at frequency 3 Hz and amplitude 20 mm generated by a swinging mechanical arm in the frontal edge of the plate [5].

One should note in this connection that Lighthill's theory of fish locomotion assumes that the amplitudes of propulsive waves created by fish body motion grow from zero on the front (at fish heads) to their maximum values at the tails. This is consistent with fish body motion in nature, but is not compatible with the behaviour of the propagating localised flexural waves employed for propulsion in the model vessels used in [3-5].

To overcome the problem of predicting zero thrust for the wave propulsive systems used in [3-5], it was suggested that nonlinear distortion of localised flexural waves may be responsible for generation of thrust in the real experimental marine craft. Indeed, the Mach numbers of propagating flexural waves used for propulsion in the experimental works [3-5] were as large as about two [3], which makes the above suggestion quite realistic. This hypothesis will be explored in the present work by adding nonlinear harmonics of propulsive flexural waves, that are growing with the distance of propagation due to elastic nonlinearity, to the Lighthill's formula for generated thrust.

For simplicity, only the lowest (third) harmonic of the localised flexural waves is used, similarly to the earlier work [10]. Also for simplicity, the effect of wave dispersion on generation of the nonlinear harmonic is neglected, assuming that the plate-like structures used in the experiments of [3-5] are short enough in the direction of wave propagation. The sum of the initial time-harmonic wave and its third nonlinear harmonic having the amplitude linearly increasing with the distance of propagation is then substituted into Lighthill's formula to derive the analytical expression for the averaged thrust.

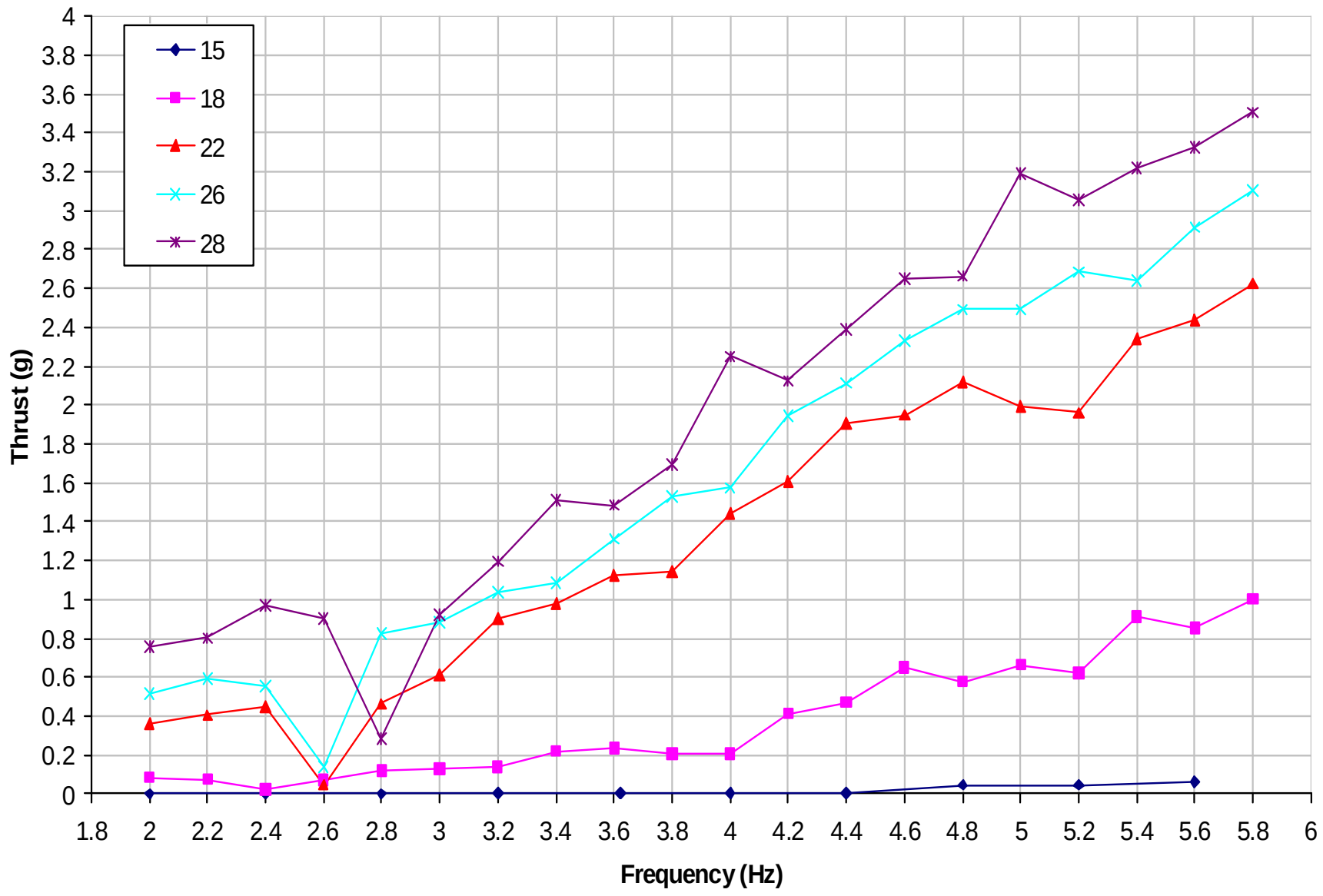

Fig. 3 Experimental results for generated thrust as functions of frequency and amplitude; the measurements have been carried out for a mono-hull model boat with a propulsive rubber plate [5].

The derived analytical expression for the averaged thrust shows that, due to the effect of the third harmonic, the thrust is no longer zero, thereby demonstrating that nonlinear distortion of the propulsive flexural waves may be important for the existence of wave-like aquatic propulsion of small marine craft employing freely propagating flexural waves [3-5]. Using the derived expression, the initial numerical estimates of the generated thrust have been carried out for the parameters of the experimental vessels and of the propulsive flexural waves used in the experiments.

## 2 Theoretical Background

### 2.1 Lighthill's approach to the theory of wave-like aquatic propulsion

One of the first theoretical papers on wave-like aquatic propulsion in application to fish locomotion has been published by Lighthill [8]. In his analysis of the problem, Lighthill considered a fish that remains stationary in a steady flow of water with the velocity $U$ in the *x*-direction. It is assumed that when the fish is motionless, or ‘stretched straight’, there is no normal force acting upon the cross section. It was also assumed that the motion of the fish at any particular cross section can be modelled as a displacement $h$ in the perpendicular direction (*z*-direction), which is a function of $x$ and $t$. This displacement causes the velocity of the fluid flowing past the cross section to change from the initial value of $U$ to a new value, $V$, which is also a function of distance and time:

$$V(x,t) = \frac{\partial h}{\partial t} + U\frac{\partial h}{\partial x}. \tag{1}$$

The mean mechanical work over a long time $\overline{W}$ done by the fish by making displacements $h(x, t)$ can be expressed in the form [8]:

$$\overline{W} = \rho U \left[ \overline{\frac{\partial h}{\partial t} V A(x)} \right]_0^L , \tag{2}$$

where $A(x)$ is the area of the circumscribing circle of the ellipse-shaped cross section of the fish. Lighthill then assumes that the function $A(x)$ and/or $h(x, t)$ has a zero value for $x = 0$ (at fish head). He also assumes that $h(x, t)$ grows towards the tail to reach its maximum there (at $x = L$), which is a good approximation for real fish body motion in nature. In these cases Eqn (2) reduces to

$$\overline{W} = \rho U \left[ \overline{\frac{\partial h}{\partial t} V A(x)} \right]_{x=L} , \tag{3}$$

which results in the following expression for the total thrust produced [8]:

$$\overline{P} = \frac{1}{2}\rho A(L) \left\{ \overline{\left(\frac{\partial h}{\partial t}\right)^2} - U^2 \overline{\left(\frac{\partial h}{\partial x}\right)^2} \right\}_{x=L} . \tag{4}$$

It is natural to try to apply the above expressions to the above-mentioned model marine craft investigated in [3-5], keeping in mind that localised flexural waves used for propulsion in [3-5] were excited from the front edges of the propulsive plates via mechanical arms, which means that, contrary to Lighthill's assumption of $h(x,t) = 0$ at $x = 0$, the amplitudes of these waves were maximal at the fronts of the propulsive plates (at $x = 0$) and were decreasing (or at best were kept constant) over the length of the plates towards the rear edges. Also, instead of $A(x) = 0$ at $x = 0$, there is $A(x) = A = const \neq 0$.

Let us now assume that the flexural waves that were generated in the experiments of the papers [3-5] are time-harmonic and have a constant amplitude $H$ along the length of the flexible fin:

$$h(x,t) = H\cos(\omega t - kx). \tag{5}$$

Here $k = \omega/c$ is the wavenumber of the localised flexural wave, where $\omega$ is the circular frequency, and $c$ is the velocity of the localised wave propagation along the propulsive plate. It must be noted that in the case of localised flexural waves used in the experiments [3-5] the wave amplitudes were not constant at different points along the perpendicular direction (y-axis). However, for simplicity, it is assumed in Eqn (5) that the amplitudes are constant everywhere.

As it follows from Eqn (5), the Lighthill's assumption of $h(x, t) = 0$ at $x = 0$ is no longer applicable, and in order to calculate the mean mechanical work $W$ one should use Eqn (2) instead of Eqn (3). Similarly, instead of using Eqn (4) to calculate thrust, one should use the full expression

$$\overline{P} = \frac{1}{2}\rho A\left\{\overline{\left(\frac{\partial h}{\partial t}\right)^2} - U^2\overline{\left(\frac{\partial h}{\partial x}\right)^2}\right\}_0^L . \quad (6)$$

Here $A = \pi(d^2/4)$, where $d$ is the width of the propulsive plate.

Substituting Eqn (5) into Eqn (6), one can see that the resulting long-time average thrust is zero. This means that there is no thrust produced by a propulsive wave of constant amplitude described by Eqn (5). However, this contradicts the experimental results [3-5] showing that there is a significant amount of thrust generated. Therefore, one can conclude that the above-mentioned theoretical analysis must be neglecting some mechanisms that are present in real experiments. It is suggested in the present paper that one of such missing mechanisms may be nonlinear waveform distortion [11, 12] of the propulsive waves during their propagation from the front to the rear edge of the plate. A detailed exploration of this idea is presented in the next section.

## 2.2 Accounting for generation of nonlinear harmonics

A possible explanation for the difference in results between the Lighthill's theory in the form of Eqn (6) and the experiments [3-5] may be nonlinear distortion of the waveforms of the propulsive waves during their propagation that may result in generation of a non-zero thrust. Using a spectral interpretation, this nonlinear distortion can be described as occurrence of higher order harmonic waves generated along the length of the propulsive plate. This hypothesis is backed up by the experimental data from the paper [5], as the results for the lowest wave amplitude tested (15 mm displacement) were practically zero, and only at the larger amplitudes there was a significant thrust achieved (see Fig. 3).

In the earlier published theoretical paper on generation of nonlinear harmonics in anti-symmetric wedge elastic waves [10], it was shown that in the case of anti-symmetric localised waves, which is also the case for the propulsive waves used in the experiments [3-5], the lowest order of nonlinearity is the third order, as the quadratic term vanishes because of the symmetry of the problem. Therefore, the description of the nonlinear distortion of the time-harmonic propulsive waves for the problem under consideration can be limited to the accounting for the third nonlinear harmonic only, for simplicity.

In light of the above, let us consider the addition of a nonlinear third harmonic to the Lighthill's formula, Eqn (6). Based on the results of [10], we will assume that the amplitude of the generated third harmonic is proportional to $H^3$, and it grows with distance $x$ linearly, starting from zero at the front edge of the propulsive plate. The addition of the third harmonic thus changes the expression (5) for $h(x,t)$, which now takes the form:

$$h(x,t) = H\cos(\omega t - kx) + F_{nl}(H)\,x\cos(3\omega t - 3kx - \psi). \quad (7)$$

Here $F_{nl}(H)$, which is proportional to $H^3$, is a non-dimensional function describing the effect of nonlinearity, and $\psi$ is the initial phase. The function $F_{nl}(H)$ also depends on the nonlinear elastic moduli of the material of the propulsive plate, which constitutes 'elastic nonlinearity'. In what follows we will assume that the material of the propulsive plate is rubber, as it was in the experiments [3-5].

In the expression (7), it is assumed that $H = const$, i.e. that the amplitude of the first (main) harmonic does not change with the distance as a result of nonlinear distortion. This initial rather rough approximation, which can be called the 'approximation of a given field' [11], will be considered here first. Later on, we will take the change of amplitude of the first harmonics into account using energy conservation law.

Substituting Eqn (7) into Eqn (6) and doing the required operations, we obtain that the generated thrust is no longer zero, and it is defined by the following expression:

$$\overline{P} = \frac{9}{4}\rho A F_{nl}{}^2(H) L^2 \omega^2 \left(1 - \frac{U^2}{c^2}\right). \quad (8)$$

It follows from Eqn (8) that thrust is generated due to the nonlinearity if the expression in brackets is positive, i.e. when $c > U$, which is the usual condition for the velocity $c$ of an elastic wave propagating along the plate to be larger than the velocity of swimming $U$. Although the expression (8) for the thrust is meaningful, the correctness of the assumption $H = const$ in application to this problem does not look very convincing.

For that reason, we now consider a more refined approach taking into account the change in the amplitude of the first harmonic with the distance of propagation because of the nonlinear generation of the third harmonic. This can be done using energy conservation law. In the approximation of only two interacting harmonics, the first and the third, this law takes the form

$$\omega^2 H^2 = \omega^2 H^2(x) + (3\omega)^2 H_3{}^2(x) = \omega^2 H^2(x) + 9\omega^2 F_{nl}{}^2(H) x^2, \quad (9)$$

where $H$ now denotes the initial amplitude of the first harmonics (at $x = 0$), and $H(x)$ represents its changing value at $x > 0$. It follows from Eqn (9) that the changing amplitude $H(x)$ can be expressed as

$$H(x) = \sqrt{H^2 - 9F_{nl}{}^2(H)x^2}\,. \quad (10)$$

We now expand the expression in the right-hand side of Eqn (10) into the Taylor series, retaining terms up to the fourth order in $F_{nl}$. This gives the following approximate expression for $H(x)$:

$$H(x) = H - \frac{9F_{nl}{}^2(H)x^2}{2H} + \frac{81F_{nl}{}^4(H)x^4}{8H^3}. \quad (11)$$

Let us now replace Eqn (7) with the more precise expression taking into account Eqn (11),

$$h(x,t)=\left(H-\frac{9F_{nl}{}^{2}(H)x^{2}}{2H}+\frac{81F_{nl}{}^{4}(H)x^{4}}{8H^{3}}\right)\cos(\omega t-kx)+F_{nl}(H)\,x\cos(3\omega t-3kx-\psi),$$

(12)

and substitute it into Eqn (6) for generated thrust, keeping terms up to the fourth power of $F_{nl}(H)$. After rather bulky derivations, it can be shown that all terms of the second order in $F_{nl}(H)$ cancel each other, and the resulting expression for generated thrust takes the form

$$\overline{P}=\frac{1}{2}\rho A\frac{81}{4}\frac{F_{nl}{}^{4}(H)\omega^{2}L^{4}}{H^{2}}\left[1-\frac{U^{2}}{c^{2}}\left(1+\frac{2}{k^{2}L^{2}}\right)\right], \tag{13}$$

where $k = \omega/c$ is the wavenumber of the propagating flexural wave. This means that the initial ‘approximation of a given field’, Eqn (7), is insufficient, and the resulting expression for generated thrust, Eqn (8), is incorrect as it does not take into account some terms of the second order in $F_{nl}(H)$ appearing due to the change of amplitude of the first harmonic with the propagation distance, which results in the mutual cancellation of all terms of the same order, and thus in zero thrust generated in this order of nonlinearity. Using a more refined formula for $h(x, t)$, Eqn (12), and keeping all the terms of the fourth order versus $F_{nl}(H)$ in further derivations results in the expression for generated thrust, Eqn (13), that shows its proportionality to the fourth power of $F_{nl}(H)$.

It is convenient to simplify Eqn (13) using a typical relationship between the parameters of the problem. Usually, $L \approx 2\lambda = 4\pi/k$. Therefore, $k^2L^2 \approx 16\pi^2 \approx 158$ and $2/(k^2L^2) \approx 0.013$, which is much less than $1$. Thus, the second term in round brackets can be neglected. This results in a simplified expression for generated thrust:

$$\overline{P}=\frac{1}{2}\rho A\frac{81}{4}\frac{F_{nl}{}^{4}(H)\omega^{2}L^{4}}{H^{2}}\left(1-\frac{U^{2}}{c^{2}}\right). \tag{14}$$

Let us now specify the nonlinear function $F_{nl}(H)$. The easiest way to proceed is to use the results of the work [10], where this function has been calculated for wedge elastic waves, which are localised flexural modes propagating along wedge tips (Fig. 4). Although linear elastic wedges are not exactly the structures that have been used in the experiments [3-5], they are interesting for their own sake and they also can be used for rough estimates of the experimental situation [3-5]. According to [10], the function $F_{nl}(H)$ for elastic wedges in vacuum takes the form

$$F_{nl}(H)=\frac{1}{4}\frac{P}{Q}\frac{b}{a}\frac{\theta^{2}}{n^{2}}(kH)^{3}. \tag{15}$$

Here $P$ and $Q$ are dimensionless parameters depending on modal shapes of wedge modes, $a = E/12(1-\sigma^2) = \rho_s c_p{}^2/12$ is a non-specified parameter, where $E$ is the Young's modulus, $\sigma$ is the Poisson ratio, $\rho_s$ is the mass density of the wedge material, and $c_p$ is the velocity of plate compression waves for the wedge material, $\theta$ is the wedge angle, and $k = \omega/c$ is the wavenumber of a wedge mode characterised by the number $n$, where $c$ is the velocity of wedge mode of number $n$ (for shortness, $k$ and $c$ are written without index $n$). The parameter $b = 0.4f(c_t/c_l)^6$ describes the nonlinear properties of the structure. Here $c_t$ and $c_l$ are the velocities of shear and longitudinal elastic waves in the wedge material, and $f$ is the relevant 4th order elastic module describing cubic nonlinearity.

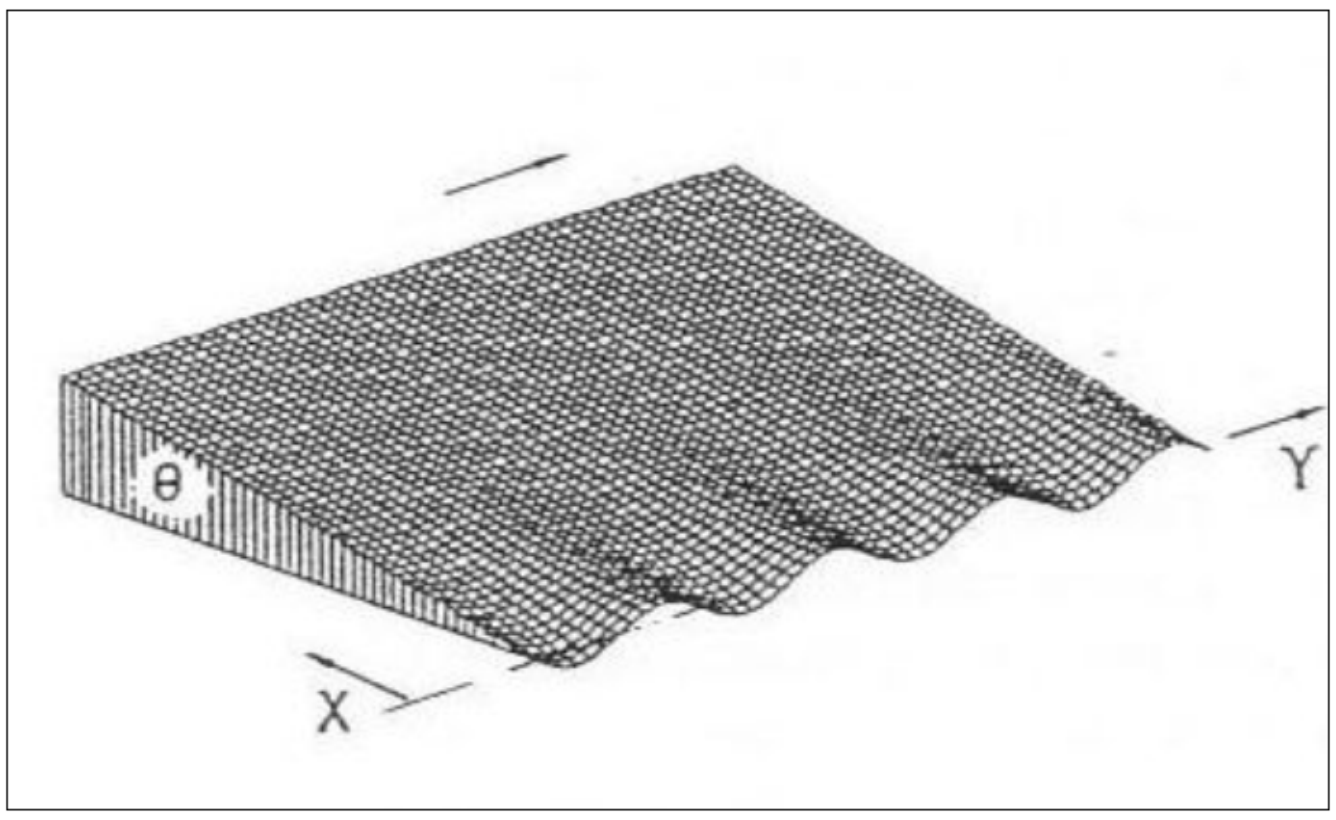

Fig. 4 Localised flexural waves propagating along the tip of a linear elastic wedge, also known as wedge elastic waves.

For elastic wedges immersed in water, we will calculate $F_{nl}(H)$ using the same Eqn (15), but with the wavenumbers $k = \omega/c$ for wedge waves in water. This means that instead of wedge wave velocities $c$ for wedges in vacuum [13, 14],

$$c = \frac{c_p}{\sqrt{3}}\theta n, \quad (16)$$

we will use the expression for wedge wave velocities for wedges in water [2],

$$c = c_t A_0^{-5/2} D^{-3/2} (\pi n)^{3/2} \theta^{3/2}, \quad (17)$$

where

$$A_0 = 6^{1/5}\left(\frac{\rho}{\rho_s}\right)^{1/5}[2(1-\sigma)]^{1/5}, \quad (18)$$

and $D = 2.102$.

## 3 Numerical Calculations and Discussion

We have carried out numerical estimates of the thrust generated by the first order ($n = 1$) localised wedge mode for a given swimming speed $U$ using formulas (14), (15) and (17), (18) for the parameters of the problem shown in Table 1. Because of the lack of reliable information about the forth order elastic moduli of rubber, we used a typical value of the relation between the forth and second order moduli, assuming that $f/E \approx 10$. For the fraction $P/Q$, we took the estimate value of 2, which was based on the numerical calculations of the paper [10].

**Table 1: Values of the parameters used in calculations**

| Parameter | Notation | Value |
|---|---|---|
| Wedge wave displacement amplitude (m) | $H$ | 0.028 |
| Wedge angle (degrees) | $\theta$ | 5 |
| Mass density of rubber (kg/m$^3$) | $\rho_s$ | 1100 |
| Shear wave speed in rubber (m/s) | $c_t$ | 30 |
| Swimming speed (m/s) | $U$ | 0.23 |
| Propulsive length (m) | $L$ | 0.25 |
| Effective width of fin (m) | $d$ | 0.055 |
| Poisson ratio of rubber | $\sigma$ | 0.49 |

For convenience of comparison with the experiments [5], the thrust was calculated in gramms, rather than in Newtons. We remind the reader that the relationship between the same forces $F$ expressed in Newtons (N) and in gramms (g) is

$$F_{(g)} = F_{(N)} \frac{g}{1000}, \qquad (19)$$

where $g = 9.81$ m/s$^2$ is gravity acceleration. The results of the calculations are shown in Fig. 5 in the frequency range 2 – 4.8 Hz. The experimental results from the paper [5] are also shown in Fig. 5 for comparison (see also Fig. 3).

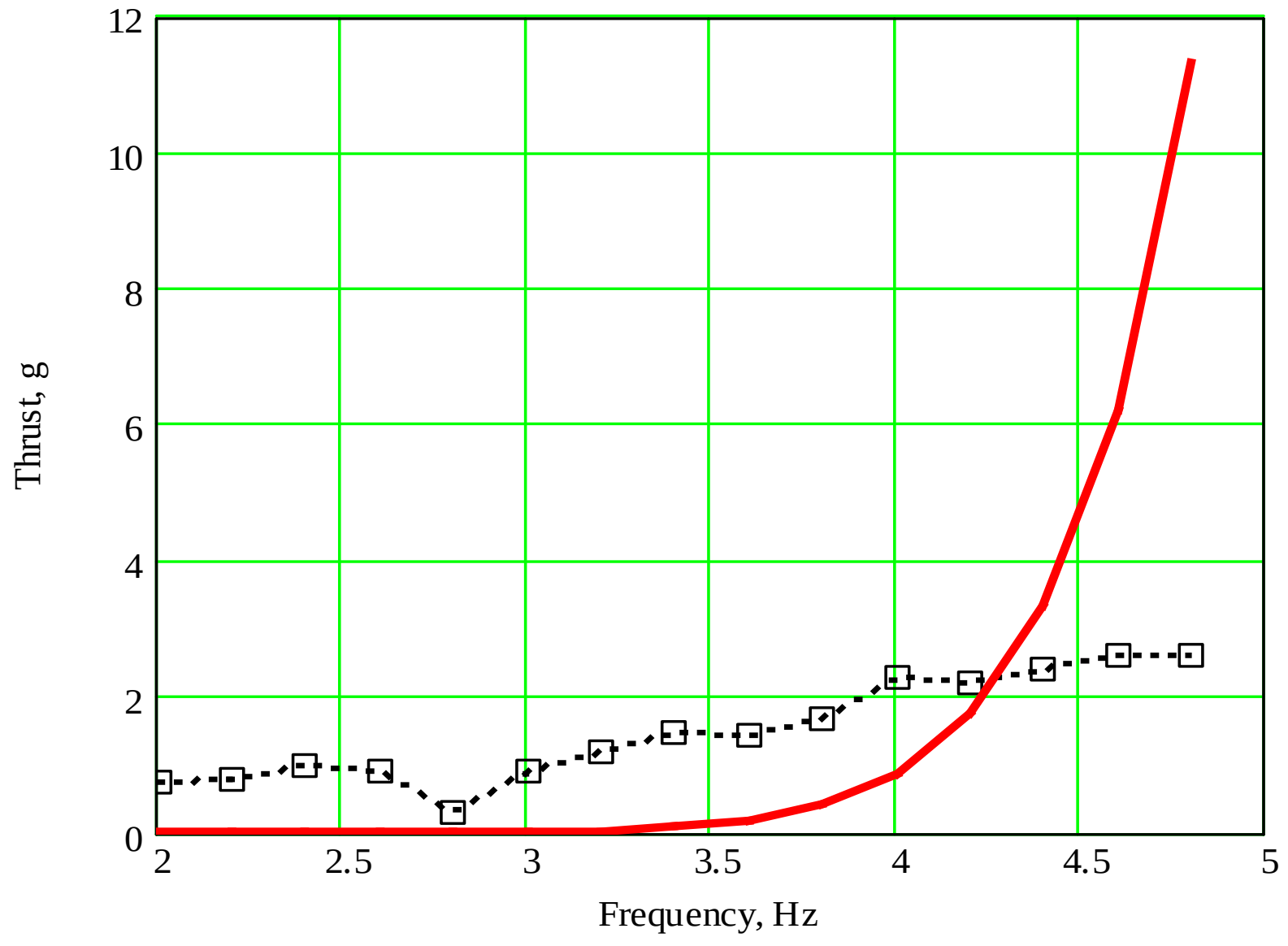

Fig. 5 Calculated thrust as function of frequency (solid line) in comparison with the corresponding experimental data (boxes) obtained in [5].

As can be seen from Fig. 5, the nonlinear mechanism gives rather large values of thrust at higher frequencies, but underestimates it at lower frequencies. This means that although the nonlinear mechanism of thrust generation is feasible, in the current stage of the theory it does not predict realistic values observed in the experiments. Further theoretical and experimental research is required in this direction.

## 4 Conclusions

The results of the initial research into the theory of aquatic propulsion by propagating flexural waves presented in this paper demonstrate that nonlinear distortion of the propagating flexural waves may be paramount for generating a nonzero thrust.

The quantitative agreement between the preliminary theoretical calculations and the experimental measurements is not satisfactory yet. The theory underestimates the measured values of generated thrust at lower frequencies and overestimates them at higher frequencies. One of the possible reasons for that could have been modelling of the real plate-like propulsive structures used in the experiments as linear elastic wedges that cause no dispersion for propagating localised flexural waves.

Further theoretical and experimental research on this topic is needed. In particular, it would be important to acquire experimental evidence of the nonlinear distortion (if any) of localised flexural waves propagating in immersed propulsive plates.